\UseRawInputEncoding
\documentclass[aps,prl,twocolumn,superscriptaddress]{revtex4-1}
\usepackage{graphicx}
\usepackage{mathrsfs}
\usepackage{bm}
\usepackage{amsmath}
\usepackage{color}
\usepackage{dcolumn}
\usepackage{epstopdf}
\usepackage{dsfont}
\usepackage{amssymb}
\usepackage{tabularx}
\usepackage{array}
\usepackage{float}
\usepackage{colordvi}

\begin{document}
\title{Interlayer ferromagnetism and insulator-metal transition in element-doped CrI${_3}$ thin films}
\author{Shiyang Sun$^{\ddag}$}
\affiliation{College of Physics and Hebei Advanced Thin Film Laboratory, Hebei Normal University, Shijiazhuang, Hebei 050024, China}
\author{Xuyan Chen$^{\ddag}$}
\affiliation{School of Gifted Young, University of Science and Technology of China, Hefei, Anhui 230026, China}
\author{Xuqi Li$^{\ddag}$}
\affiliation{College of Physics and Hebei Advanced Thin Film Laboratory, Hebei Normal University, Shijiazhuang, Hebei 050024, China}
\author{Huihui Zhang}
\affiliation{College of Physics and Hebei Advanced Thin Film Laboratory, Hebei Normal University, Shijiazhuang, Hebei 050024, China}
\author{Haidan Sang}
\affiliation{College of Physics and Hebei Advanced Thin Film Laboratory, Hebei Normal University, Shijiazhuang, Hebei 050024, China}
\author{Shifei Qi}
\email[Correspondence author:~~]{qisf@hebtu.edu.cn}
\affiliation{College of Physics and Hebei Advanced Thin Film Laboratory, Hebei Normal University, Shijiazhuang, Hebei 050024, China}
\affiliation{International Center for Quantum Design of Functional Materials, CAS Key Laboratory of Strongly-Coupled Quantum Matter Physics, and Department of Physics,  University of Science and Technology of China, Hefei, Anhui 230026, China}
\author{Zhenhua Qiao}
\email[Correspondence author:~~]{qiao@ustc.edu.cn}
\affiliation{International Center for Quantum Design of Functional Materials, CAS Key Laboratory of Strongly-Coupled Quantum Matter Physics, and Department of Physics,  University of Science and Technology of China, Hefei, Anhui 230026, China}

\begin{abstract}
  The exploration of magnetism in two-dimensional layered materials has attracted extensive research interest. For the monoclinic phase CrI${_{3}}$ with interlayer antiferromagnetism, finding a static and robust way of realizing the intrinsic interlayer ferromagnetic coupling is desirable. In this Letter, we study the electronic structure and magnetic properties of the nonmagnetic element (e.g., O, S, Se, N, P, As and C) doped bi- and triple-layer CrI$_3$ systems via first-principles calculations. Our results demonstrate that O, P, S, As, and Se doped CrI$_3$ bilayer can realize interlayer ferromagnetism. Further analysis shows that the interlayer ferromagnetic coupling in the doped few-layer CrI${_{3}}$ is closely related to the formation of localized spin-polarized state. This finding indicates that insulated interlayer ferromagnetism can be realized at high doping concentration (larger than 8.33$\%$). When the doping concentration is less than 8.33$\%$, but larger than 2.08$\%$, an insulator-metal phase transition can occur since the localized spin-polarized states percolate to form contiguous grids in few-layer CrI${_{3}}$.
\end{abstract}

\maketitle

\textit{Introduction---.} Two-dimensional(2D) magnetic semiconductors have attracted extensive attention due to the enormous potential for novel magneto-optic~\cite{Huang1,Gong,Deng,Gibertini, Zhong,Seyler}, magnetoelectronic~\cite{Zollner,Song1,Klein,Cardoso,Wang1,Ghazaryan,Wang2,Wang3}, and spintronic devices~\cite{Cummings,Kim1,Karpiak}. As a representative 2D layered material, CrI${_{3}}$ possesses its own unique physical properties. The bulk CrI${_{3}}$ has two different structures, i.e., high-temperature monoclinic phase and low-temperature rhombohedral phase. The bilayer CrI${_{3}}$ with rhombohedral stacking exhibits interlayer ferromagnetic coupling, while that with monoclinic stacking exhibits interlayer antiferromagnetic coupling~\cite{Sivadas,Jiang1,Jang,Soriano1,Thiel,Ubrig,Kim2}, and their phase transition temperature is 220 K~\cite{McGuire}. Thus, the magnetic order of CrI${_{3}}$ is susceptible to the variation of layer thickness and stacking order. Previous first-principles calculations have predicted that interlayer magnetic coupling can be effectively modulated by stacking order in bilayer ~\cite{Sivadas,Jiang1,Jang}. Later experiments~\cite{Thiel,Ubrig} approved that different stacking orders can affect the observed magnetic states of CrI${_{3}}$ in both bulk and few-layer CrI${_{3}}$ systems.

In van de Waals layered systems, the relatively weak interlayer coupling indicates that the interlayer magnetic order can be easily tuned via external means. Indeed, it was experimentally reported that monoclinic bilayer CrI${_{3}}$ can be transformed from interlayer antiferromagnetic to ferromagnetic coupling by applying electric gating~\cite{Jiang2,Jiang3,Huang2,Xu}. A possible physical mechanism describing this magnetic transition is the formation of magnetic polaron, which was theoretically confirmed~\cite{Soriano2}. Beside above external electric gating, a natural question arises: whether it is possible to find a static and robust way of realizing the intrinsic interlayer ferromagnetic coupling in few-layer CrI${_{3}}$? In addition, for semiconductor materials, the carrier doping concentration may destroy the physical properties. Therefore, it is desirable to realize the interlayer ferromagnetically-coupled few-layer CrI${_{3}}$ while maintaining its semiconducting characteristics without introducing additional carriers.

In this Letter, we perform a systematic study on the magnetic and electronic properties of nonmagnetic-element doped few-layer CrI${_{3}}$ by using first-principles calculation methods. We first show that the interlayer ferromagnetic coupling can be established in bilayer CrI${_{3}}$ doped with C, N, O, P, S, As, or Se. We then find that the interlayer ferromagnetic coupling is intimately related to the formation of magnetic polaron. Especially for the As-doped bi- or tri-layer CrI${_{3}}$, it can achieve higher Curie temperature and does not introduce extra carriers in the presence of increasing doping concentration within certain scale, therefore maintaining the system's semiconducting properties. In addition, an insulator-metal phase transition occurs with the help of percolated spin-polarized states at low doping concentration.

\begin{figure}
  \centering
  \includegraphics[width=8.6cm,angle=0]{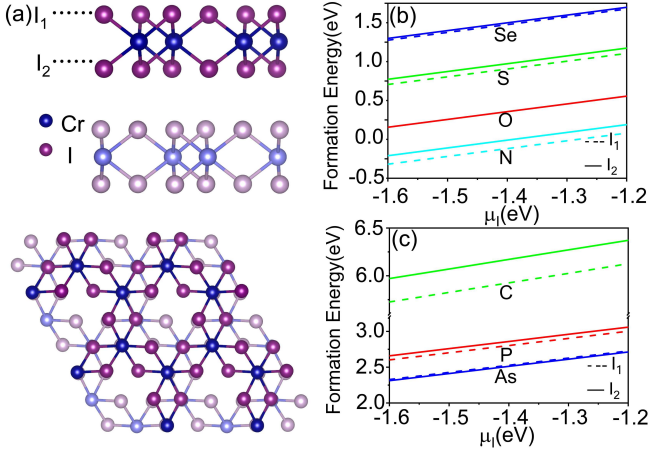}
  \caption{(a) Side and top views of crystal structures of high-temperature monoclinic bilayer CrI${_{3}}$ phase and the substitution sites are labeled as I${_{1}}$ and I${_{2}}$. Formation energies of (b) O, S, Se, N and (c) P, As or C element-doped bilayer CrI${_{3}}$ as a function of the host element chemical potentials.}
  \label{fig1}
\end{figure}

\textit{Calculation Methods---.} Our first-principle calculations were performed by using the projected augmented-wave method~\cite{Blochl} as implemented in the Vienna ab initio simulation package (VASP)~\cite{Kresse1,Kresse2}. The generalized gradient approximation (GGA) of Perdew-Burke-Ernzerhof (PBE) type was used to treat the exchange-correlation interaction~\cite{Perdew}. In our calculations, the lattice constant of the high-temperature phase of CrI${_{3}}$ was chosen to be $a_0$=6.92~\AA~\cite{Jiang1}. A vacuum buffer space of 15~\AA was used to prevent the coupling between adjacent slabs. The kinetic energy cutoff was set to be 340~eV. With fixed supercells, all structures were fully relaxed. The van der Waals (vdW) force was taken into account by employing the Grimme's method (DFT-D2)~\cite{Grimme}. The Brillouin-zone integration was carried out by using $5\times 5\times1$ Monkhorst-Pack grids. Unless mentioned otherwise, GGA+U~\cite{Anisimov,Dudarev} method was used with the on-site repulsion parameter $U=3.9~eV$ and the exchange parameter $J=1.1~eV$~\cite{Jiang1}, where $U$ is for the more localized $3d$ orbitals of Cr atoms. The Curie temperature $T{_{C}}$ was estimated within the mean-field approximation by using $k{_{B}}T{_{C}}=2/3Jx$~\cite{Bergqvist}, where $k{_{B}}$ is the Boltzmann constant, $x$ is the dopant concentration, and $J$ is the exchange parameter obtained from the total energy difference between ferromagnetic and antiferromagnetic configurations.

\begin{figure}
  \centering
  \includegraphics[width=8.6cm,angle=0]{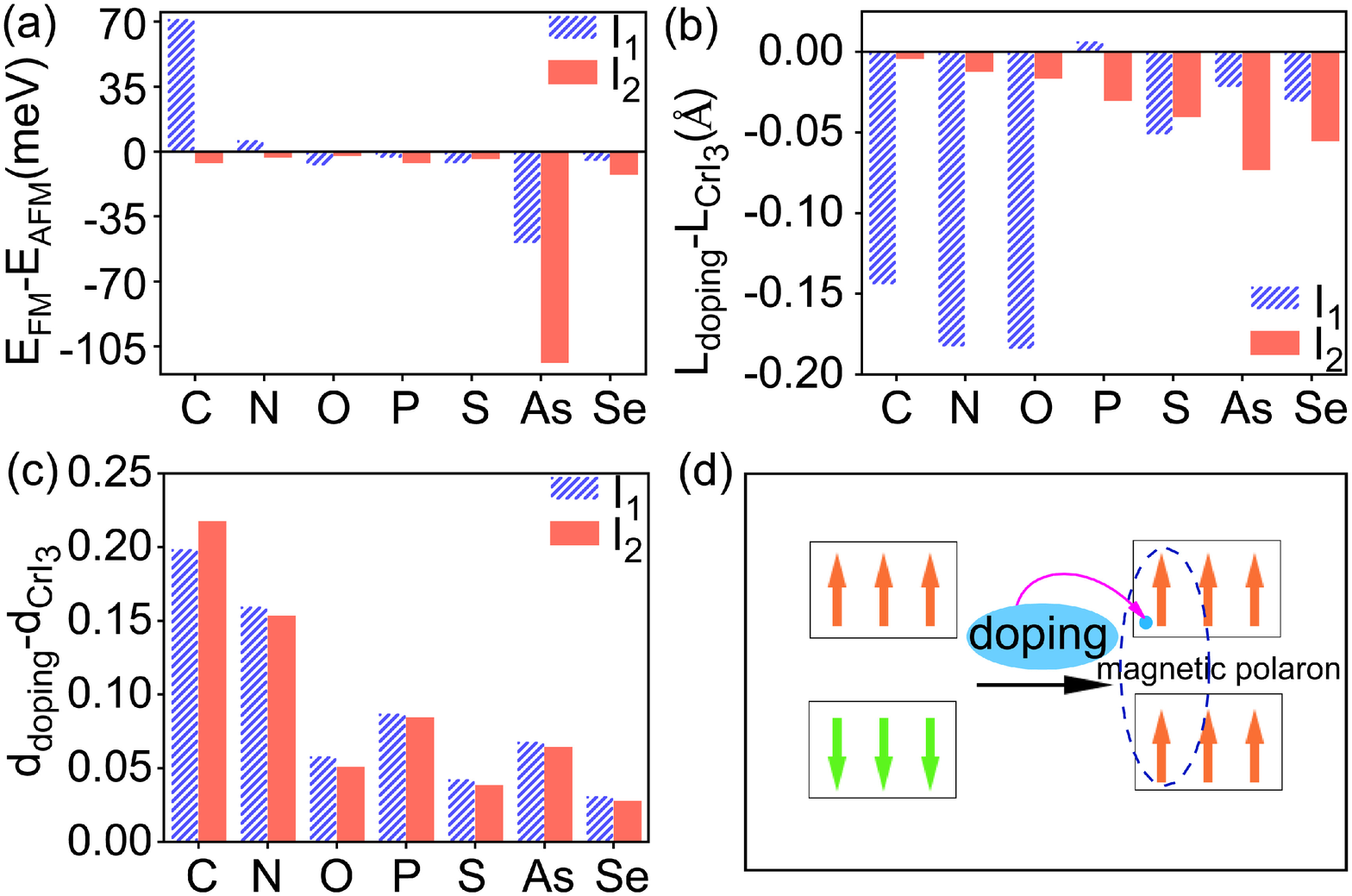}
  \caption{(a) Energy difference between interlayer ferromagnetic (FM) and antiferromagnetic (AFM) states. (b) Difference of interlayer distance  between doping configuration and pristine CrI${_{3}}$. (c) Charge difference between Cr atoms near doping site in the doped and pristine bilayer CrI${_{3}}$. (d) Schematic illustration of spin-polarized state-mediated interlayer ferromagneitic coupling in doped bilayer CrI${_{3}}$.}
  \label{fig2}
\end{figure}

\textit{Experimental Possibility of Element Doping---.} We first study the possibility of element doping in bilayer CrI${_{3}}$, i.e., substituting I by nonmagnetic dopants. Some typical candidates of nonmagnetic dopants including O, S, Se, N, P, As and C are considered. As displayed in Fig.~1(a), there are two types of I-doping sites labelled as I${_{1}}$ (at the surface) and I${_{2}}$ (inside the interlayer). The formation energy was obtained by using the expression~\cite{Zhang,Qi,Han} $\Delta H{_{F}}= E{_{tot}^{D}} -E{_{tot}}-\Sigma n{_{i}}\mu_{i}$, where $E{_{tot}^{D}}$ is the total energy of the system including one nonmagnetic impurity, $E{_{tot}}$ is the total energy of the system, ${\mu_{i}}$ is the chemical potential for the species $i$ (host atoms or dopants), and n${_{i}}$ is the corresponding number that was added/removed from the system.

As displayed in Fig.~\ref{fig1}(b), for O, S, Se, N substitutions at two I sites in the same CrI${_{3}}$ layer, the formation energy is within the range of $-0.4 \sim 1.5~eV$. It indicates that the I${_1}$ substitutional site is preferred due to smaller formation energy than that at I${_2}$ substitution. For example, N substitution leads to smaller formation energy (about $-0.6 \sim 0.2eV$) than those from O (about $-0.2 \sim 0.2eV$), S (about $0.4 \sim 0.9eV$) and Se (about $1.1 \sim 1.4eV$) substitution in the whole range of the accessible host element chemical potentials. However, for P, As, and C substitutions [see Fig.~\ref{fig1}(c)], they have larger formation energies than those with O, S, Se or N dopants. In addition, we find that the I${_{1}}$ substitutional site is preferred by P and C, while I${_{2}}$ substitutional site is preferred by As. The formation energy shows that all candidate elements (except As) are more stable at I${_{1}}$ position. The formation energy of As substituted I${_{2}}$ site is positive (about $2.3 \sim 2.7eV$). It is noteworthy that C-doped ZnO has been experimentally fabricated even the estimated formation energy is about 5.3 eV~\cite{Pan}, which is much larger than all element-doped CrI${_{3}}$. Therefore, it is reasonable to believe that O, S, Se, N, P and As doped CrI${_{3}}$ bilayer could be experimentally fabricated.

\textit{Magnetic Properties---.} We now move to investigate the interlayer magnetic coupling of the doped CrI${_{3}}$. Figure~\ref{fig2}(a) displays the energy difference $\Delta$${E}$=${E_{FM}}$-${E_{AFM}}$ between interlayer ferromagnetic and antiferromagnetic states for different element-doped bilayer CrI${_{3}}$. As reported, the pristine bilayer CrI${_{3}}$ exhibits interlayer antiferromagnetic coupling~\cite{Sivadas,Jiang1,Jang}. The introduction of dopants except C and N leads to $\Delta E<0$, indicating the formation of interlayer ferromagnetism. For the O, P, S, As, and Se substitution at I${_{1}}$(I${_{2}}$) site [see Fig.~\ref{fig2}(a)], $\Delta E$ are respectively -7.7(-2.2), -3.6(-6.3), -6.7(-4.0), -49.7(-113.9), and -5.4(-12.3) meV, indicating the interlayer ferromagnetic coupling. As a contrast, it maintains the interlayer antiferromagnetism in C or N doped case. Hereinbelow, we choose the I${_{2}}$-site As-doped bilayer CrI${_{3}}$ as an example to analyze the origin of the interlayer ferromagnetism.

\begin{figure}
  \centering
  \includegraphics[width=8.6cm,angle=0]{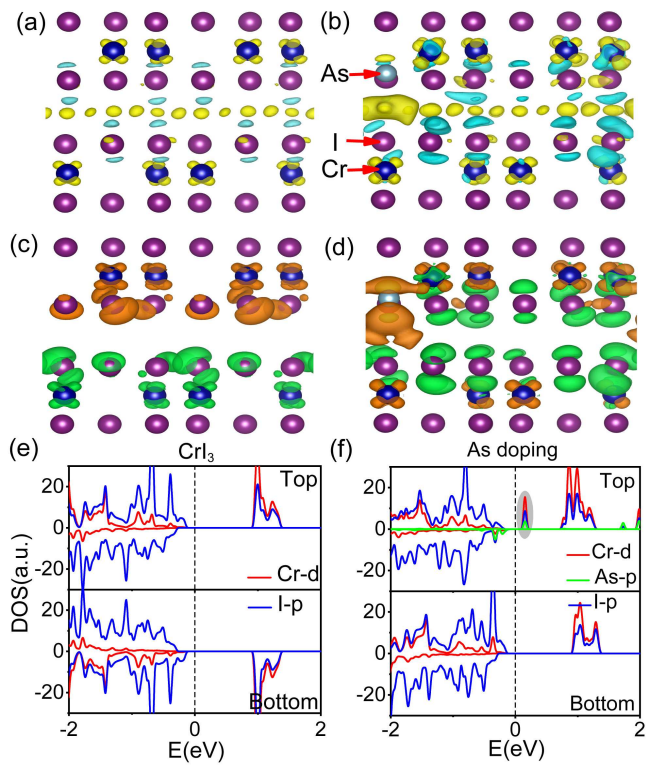}
  \caption{Differential charge density of (a) pristine and (b) As-doped bilayer CrI${_{3}}$. Spin density of (c) pristine and (d) As-doped bilayer CrI${_{3}}$. Local density of states of (e) pristine and (f) As-doped bilayer CrI${_{3}}$. Yellow and blue isosurfaces represent respectively charge accumulation and reduction. Red and green isosurfaces represent respectively spin up and spin down. Cr-d, I-p and As-p orbitals in each layer of CrI${_{3}}$ are displayed.}
  \label{fig3}
\end{figure}

For vdW magnetic materials, many studies have shown that the interlayer distance plays a crucial role in determining the interlayer magnetic coupling~\cite{Li1,Song2,Xia,Zhu1}. Thus, we first investigate the relationship between the interlayer distance and the energy difference $\Delta E$. Figure~\ref{fig2}(b) displays the difference of interlayer distances between doped and pristine CrI${_{3}}$. One can find that the interlayer distances in nearly all doped systems [except P-doped configuration at I${_{1}}$ substitution)] decrease with respect to the pristine case. Particularly, the interlayer distances in C, N, and O doped systems at I${_{1}}$-site substitution shrink respectively about 0.14, 0.17, and 0.18~\AA, which are much larger than that in the As-doped system. These together show that there is no obvious correlation between the strength of ferromagnetic coupling and the interlayer distance.

It was reported that the interlayer magnetism of vdW materials is closely related to different 3d electron occupation between different layers~\cite{Xiao,Li2,Zhu2}. In Fig.~\ref{fig2}(c), we display the charge difference of Cr atoms near the dopants between the doped and pristine bilayer CrI${_{3}}$. It shows that the charge of Cr atoms near dopants increases for all doped systems. However, in As-doped bilayer, the change of 3d electron occupation due to doping is much less than those of the C and N doped bilayers. Therefore, the difference of 3d electrons occupation between two layers cannot explain the formation of the interlayer ferromagnetism.

\begin{figure*}
  \centering
  \includegraphics[width=18.0cm,angle=0]{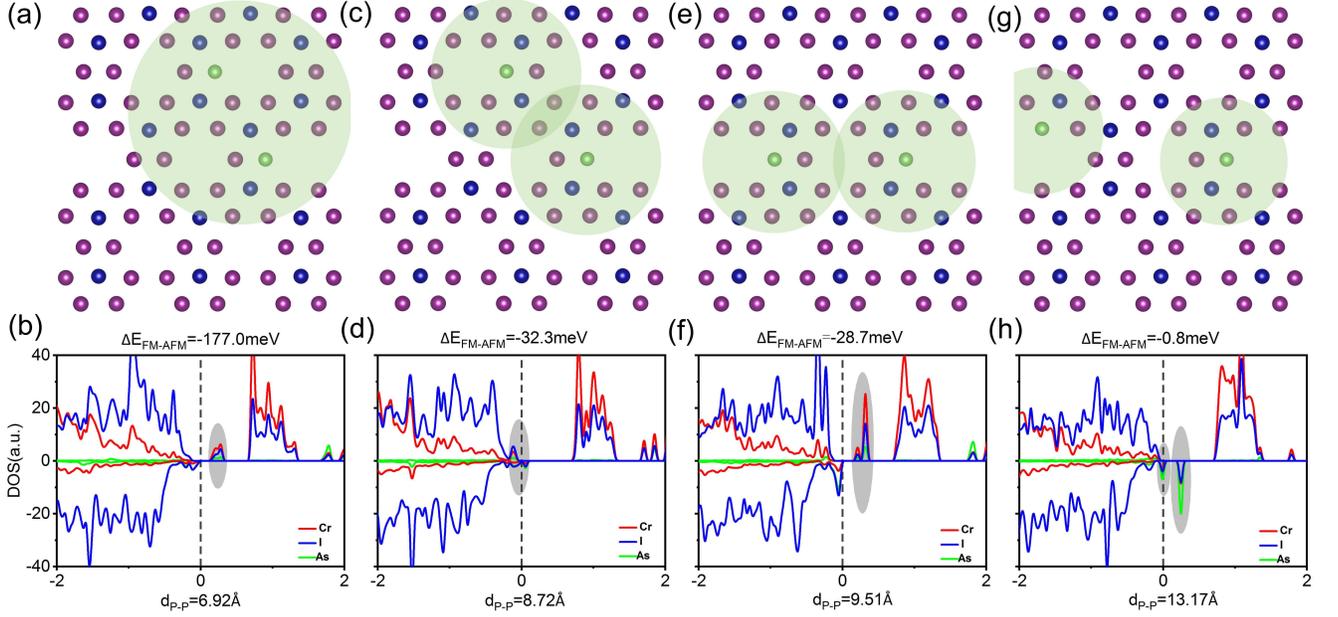}
  \caption{(a-c-e-g) Schematic plots of two spin-polarized states at different distances (d${_{P-P}}$) in As-doped bilayer CrI${_{3}}$. (b-d-f-h) The density of states of ferromagnetic state and the energy difference between interlayer ferromagnetic (FM) and antiferromagnetic (AFM) states at different d${_{P-P}}$ in As-doped bilayer. The shadow part indicates the formation of spin-polarized state.}
  \label{fig4}
\end{figure*}

We now move to calculate the differential charge densities of pristine and As-doped systems [see Figs.~\ref{fig3}(a) and \ref{fig3}(b)]~\cite{Jiang1,Liu}. One can see that As-doping indeed leads to obvious change of charge distribution inside the interlayer space. Figures~\ref{fig3}(c) and~\ref{fig3}(d) display the spin densities of pristine and doped cases, respectively. For pristine case, it exhibits intralayer ferromagnetism and interlayer anti-ferromagnetism. After As-doping, the interlayer anti-ferromagnetism transits to ferromagnetism, accompanying with a strongly localized spin-polarized state near the doping site. To confirm this finding, the local density of states of pristine and As doped bilayers are displayed in Figs.~\ref{fig3}(e) and~\ref{fig3}(f). In Fig.~\ref{fig3}(f), one can find that an extremely sharp local density of state appears near the Fermi level, arising from the hybridization among As p-orbital, bonded Cr d-orbital in the same layer, and I p-orbital in the other layer. All together indicate that As-doping results in a bound spin-polarized state, potentially forming the interlayer ferromagnetism.

To further confirm this physical origin, in Fig.~S1, we display the density of states of the C, N, O, P, S, and Se doped bilayers, where C and N doped systems exhibit interlayer antiferromagnetism, but O, P, S, and Se doped systems exhibit interlayer ferromagnetism~\cite{SM}. Figures~S1(a) and S1(b) display that no spin-polarized bound states appear near the Fermi level for C and N doped systems. While for other doped systems, spin-polarized bound states arise near the Fermi level [see Figs.~S1(c)-S1(f)]. These suggest a direct evidence that the interlayer ferromagnetic coupling in doped bilayer CrI${_{3}}$ is a consequence of the formation of spin-polarized bound state.

Another striking transport phenomenon is the insulating nature after doping. It is known that doping or gating can result in the ferromagnetism of semiconductors, but may also break the semiconducting property due to the carrier injection. Surprisingly, for O, P, S, As, and Se doped bilayer CrI${_{3}}$, our results show that they exhibit both ferromagnetic and insulating features. In association with the experimental finding that different gate doping levels do not lead to n- or p-type conduction of bilayer CrI${_{3}}$ dominantly with affected magnetic properties~\cite{Huang2}, it is believed that only insulated interlayer ferromagnetism in few-layer CrI${_{3}}$ can be observed due to the formation of spin-polarized bound state at certain doping concentrations.

When localized spin-polarized states percolate to form contiguous grids, an insulator-metal transition can occur at certain doping concentration. As shown in Figs.~\ref{fig4}(a-c-e-g), we study the electronic structure and magnetic property of two-As-atom doped bilayer CrI${_{3}}$ at different As-As distances (i.e., various doping concentrations). For the nearest As-As distance d${_{P-P}}$=6.92~\AA, in comparison with density of states of only one-As-atom doped bilayer CrI${_{3}}$ (see Fig.~\ref{fig3}(f)), the spin-polarized states in Fig.~\ref{fig4}(b) is more delocalized and only one spin-polarized states is formed [A more larger spin-polarized states is schematically plotted in Fig.~\ref{fig4}(a)]. At this configuration (the estimated doping concentration is about 8.33$\%$), it is a ferromagnetic insulator and the interlayer ferromagnetic coupling ($\Delta E$= -177.0 meV) becomes stronger than that($\Delta E$= -113.9 meV) with only one-As dopant. When the As doping concentration decreases, from Figs.~\ref{fig4}(c) and \ref{fig4}(e), we find that two spin-polarized states can be formed and percolated. In such percolated systems, they are p-type semiconductor with interlayer ferromagnetism (see Figs.~\ref{fig4}(d) and \ref{fig4}(f)), but the ferromagnetic coupling becomes weaker than that of the insulating case. For longer As-As distance, two independent spin-polarized states are formed and the interlayer magnetic coupling almost disappears. For example, when the As-As distance d${_{P-P}}$ is 13.17~\AA, one spin-polarized state is about 6.6~\AA. If the doping concentration is further decreased to 2.08$\%$, our results indicate that an insulated and ferromagnetic bilayer CrI${_{3}}$ can be formed again due to the independent spin-polarized states, similar to that in Fig.~\ref{fig3}(f).

So far, we have shown that element doping in bilayer CrI${_{3}}$ can induce interlayer ferromagnetism. The trilayer CrI${_{3}}$ has weak interlayer ferromagnetic coupling~\cite{Huang1}. Whether the element doping can enhance the interlayer ferromagnetism? By taking As-doping as an example in Fig.~S2(a), three I substituted sites including I1, I2, and I3 are selected. The interlayer magnetic couplings and their relative stabilities are respectively displayed in Fig.~S2(b) and Table S1. One can see that the state with interlayer ferromagnetic coupling is more stable than the other three states with interlayer antiferromagnetic coupling, which agrees well with the experimental observation~\cite{Huang1}. One can find that the total energy of As substitution at I3 site is the lowest, indicating the most stable structure. Therefore, we show that the strong interlayer ferromagnetic coupling can also be established in trilayer CrI${_{3}}$ via As doping.

\textit{Conclusions---.} In summary, we demonstrate that the interlayer ferromagnetic coupling can be realized in both bilayer and trilayer CrI${_{3}}$ by doping nonmagnetic elements. Our finding provides a new evidence that the interlayer ferromagnetic coupling in CrI${_{3}}$ thin films may be related to the formation of spin-polarized states, and also provides an alternative scheme for the realization of CrI${_{3}}$ based ferromagnetic insulator.

\textit{Acknowledgements---.} This work was financially supported by the NNSFC (11974098, 11974327, and 12004369), Natural Science Foundation of Hebei Province(A2019205037), and Science Foundation of Hebei Normal University (2019B08), Fundamental Research Funds for the Central Universities (WK2030020032 and WK2340000082), Anhui Initiative in Quantum Information Technologies. The supercomputing services of AM-HPC and USTC are gratefully acknowledged.

$^{\ddag}$ These authors contribute equally to this work.

\end{document}